\documentclass[english,aps,tightenlines,showpacs, showkeys, notitlepage]{revtex4-1}
\usepackage[latin9]{inputenc}
\usepackage{amsmath}
\usepackage{amssymb}

\makeatletter

\usepackage{dcolumn}
\usepackage{bm}

\makeatother

\usepackage{babel}
\begin{document}
\title{Gauge theory of the Maxwell and semi-simple extended (anti) de Sitter
algebra}
\author{Salih Kibaro\u{g}lu}
\email{salihkibaroglu@gmail.com}

\author{Oktay Cebecio\u{g}lu}
\email{ocebecioglu@kocaeli.edu.tr}

\date{\today}
\begin{abstract}
In this paper, a semi-simple and Maxwell extension of the (anti) de
Sitter algebra is constructed. Then, a gauge-invariant model has been
presented by gauging the Maxwell semi-simple extension of the (anti)
de Sitter algebra. We firstly construct a Stelle-West like model action
for five-dimensional space-time in which the effects of spontaneous
symmetry breaking have been taken into account. In doing so, we get
an extended version of Einstein's field equations. Next, we decompose
the five-dimensional extended Lie algebra and establish a MacDowell-Mansouri
like action that contains the Einstein-Hilbert term, the cosmological
term as well as new terms coming from Maxwell extension in four-dimensional
space-time where the torsion-free condition is assumed. Finally, we
have shown that both models are equivalent for an appropriately chosen
gauge condition.
\end{abstract}
\affiliation{Department of Physics, Kocaeli University, 41380 Kocaeli, Turkey}
\maketitle

\section{Introduction}

The gauge theories provide a useful theoretical background to describe
the fundamental interactions of particle physics. The three basic
interactions, strong, weak and electromagnetic are described by Yang-Mills
gauge theory, which is based on the Lie-groups $\mathcal{SU}\left(3\right)$,
$\mathcal{SU}\left(2\right)$ and $\mathcal{U}\left(1\right)$, respectively.
These interactions form the basis of the Standard Model. The fourth
interaction, gravitational interaction, characterized by Einstein's
general theory of relativity is not a genuine gauge theory in the
sense of Yang-Mill's type. On the other hand, it can be constructed
as the Yang-Mills gauge theory by a symmetry-breaking mechanism. Before
we start, it is important to point out the non-commutativity of translation
generators, $\left[P_{a},P_{b}\right]=\pm iM_{ab}$, for the de Sitter
($dS$) groups. Stelle and West demonstrated that the Einstein-Cartan
theory is reproduced by spontaneously broken $dS$ group to the Lorentz
group \citep{stelle1979,stelle1980}. This approach opened new doors
to unify the fundamental interactions in the framework of gauge theory
\citep{vervimp1994}. 

The (anti) de Sitter $\left(A\right)dS$ gauge theory of gravity was
first formulated in papers \citep{stelle1980,townsend1977,macDowell1977,smrz1980}.
There are two main advantages of $\left(A\right)dS$ gauge theory:
The first one is to construct a consistent and renormalizable theory.
It is well-known that the general theory of relativity is not a renormalizable
theory. To overcome this problem, many theoretical models have been
proposed. Among these theories, $\left(A\right)dS$ gauge theory emerges
as a useful candidate to solve this issue because it can be constructed
as a Yang-Mills gauge theory \citep{Piguet1995}. The second one is
the explanation of the cosmological constant. The astronomical observations
show that our universe is accelerating as in de Sitter spacetime.
Therefore, one can say that $\left(A\right)dS$ gauge theory of gravity
have an important potential to explain the accelerating expansion
with a positive cosmological constant (for more detail see \citep{huang2008}). 

Similar to the de Sitter group, there exists a group having non-commuting
translational generators, called the Maxwell group. This group can
be interpreted as a modification of the Poincaré group which describes
the empty Minkowski space-time, and the translation generators are
no longer abelian but satisfy $[P_{a},P_{b}]=iZ_{ab}$ \citep{soroka2005,gomis2009,bonanos2009,bonanos2010,bonanos2010A,bonanos2010B}.
Here, the antisymmetric central charges $Z_{ab}$ represent additional
degrees of freedom. In early studies, these extra degrees of freedom
were associated with constant electromagnetic fields \citep{bacry1970,schrader1972}.
Recently, the Maxwell group and its modifications have extensively
studied to generalize Einstein\textquoteright s theory of gravity
and supergravity \citep{azcarraga2011,soroka2012,durka2012,Azcarraga2014,cebecioglu2014,cebecioglu2015,salgado2014,concha2015A,penafiel2018,ravera2018,kibaroglu2019A,kibaroglu2019B,kibaroglu2020A,concha2019A,concha2019B}.
Furthermore, gauge theory of gravity based on the Maxwell (super)algebras
lead to the cosmological constant as in $\left(A\right)dS$-like gravity
theories \citep{azcarraga2011,soroka2012,durka2012,cebecioglu2014,concha2015A,penafiel2018,kibaroglu2019B,kibaroglu2020A}.
This feature may take an essential role because the studies on the
cosmological constant indicate that there should be a background field
filling our space-time. In this new context, the additional degrees
of freedom represent the uniform gauge field strengths in (super)space
leading to uniform constant energy density \citep{bonanos2010A}.
Therefore we can say that the Maxwell symmetries may provide a powerful
geometrical framework for the cosmological constant and dark energy
\citep{freund1974,bramson1974}. So, it can be said that Maxwell symmetry
contains a useful structure to generalize the general theory of relativity.
Moreover, this symmetry is used in various areas such as describing
planar dynamics of the Landau problem \citep{fedoruk2012}, the higher
spin fields \citep{fedoruk2013,fedoruk2013A}, and it was also used
in the string theory as an internal symmetry of the matter gauge fields
\citep{hoseinzadeh2015}.

In the present paper, we consider a Maxwell generalization of $\left(A\right)dS$
gauge theory of gravity for two reasons. The first one is the compatibility
with renormalizability requirements of $dS$ gauge group. Namely,
the semi-simple structure of $dS$ gauge group makes it possible to
form the $dS$ gravity as a Yang-Mills gauge theory \citep{sobreiro2012}.
The second is the present accelerated expansion of our universe due
to the basic properties of the de Sitter geometry and its close connection
with the cosmological constant. The second is the present accelerated
expansion of our universe due to the basic properties of the de Sitter
geometry and its close connection with the cosmological constant and
it is also well known that the de Sitter vacuum keeps the position
of the best candidate for a dark matter. For these reasons, the extension
of (anti) de Sitter algebra may provide new insights into the nature
of these links. 

The organization of this letter is as follows. In Sect. II, we give
a short introduction to the Stelle-West model for gravity based on
$\left(A\right)dS$ algebra. We then give a semi-simple extension
of $\left(A\right)dS$ algebra by considering the Maxwell symmetry.
Using the resulting algebra, we construct a gauge theory of gravity
and after symmetry-breaking, we get an extended version of the Einstein
field equations. In Sect. III, firstly, we will be interested in decomposing
extended five-dimensional algebra with respect to one of its regular,
four-dimensional subalgebras. A regular subalgebra here is given by
de Sitter algebra. Secondly, we construct gauge theory of gravity
based on this decomposed algebra and establish a MacDowell-Mansouri
like action. Then, we are able to show that both actions, Stelle-West
and MacDowel Mansouri, coincide with each other in a certain gauge.

\section{The (anti) de sitter gravity and its maxwell extension}

We introduce our discussion by reviewing the theory of spontaneously
broken $\left(A\right)dS$ gravity given by the Stelle-West model
\citep{stelle1979,stelle1980,randono2010}. This model provides a
useful background for both $dS$ and $AdS$ groups. The (anti) de
Sitter algebra in 4D is
\begin{equation}
\left[\mathcal{M}_{AB},\mathcal{\mathcal{M}}_{CD}\right]=i\left(\eta_{AD}\mathcal{\mathcal{M}}_{BC}+\eta_{BC}\mathcal{\mathcal{M}}_{AD}-\eta_{AC}\mathcal{\mathcal{M}}_{BD}-\eta_{BD}\mathcal{\mathcal{M}}_{AC}\right),
\end{equation}
where $\mathcal{M}_{AB}$ are the generators of the group and the
metric tensor is chosen to be $\eta_{AB}=diag\left(1,-1,-1,-1,-1\right)$.
Here the capital Latin indices run $A,B,...=0,1,2,3,5$. 

To construct a gravitational theory based on the given symmetry group,
we will give only a summary of $\left(A\right)dS$ gravity as presented
by Stelle and West. The Stelle-West action is given by the help of
(anti) de Sitter curvature $\mathcal{F}^{AB}\left(x\right)$ and its
connection $\mathcal{A}^{AB}\left(x\right)$
\begin{equation}
S_{SW}=\sigma\int V^{E}\epsilon_{ABCDE}\mathcal{F}^{AB}\wedge\mathcal{F}^{CD}+\alpha\left(c^{2}-V_{A}V^{A}\right)\label{eq: action sw}
\end{equation}
where $V^{A}$ is a non-dynamical five vector field has a positive
constant magnitude $V_{A}V^{A}=c^{2}$. We also note that $V^{A}$
does not have any degrees of freedom, but it helps to construct the
geometrical structure of the theory \citep{stelle1979}. Moreover,
$\sigma$ is an arbitrary constant and $\alpha$ represents the Lagrange
multiplier \citep{leclerc2006,randono2010}. This action reduces to
the Einstein-Cartan action under the following constraints,

\begin{equation}
V^{A}=\left(0,0,0,0,c\right)
\end{equation}

\begin{equation}
e^{a}=-lDV^{a}=-lc\mathcal{A}{}_{\,\,5}^{a},\,\,\,\,\,\,\,\,\,DV^{5}=0\label{eq: constaints}
\end{equation}
where $e^{a}\left(x\right)$ corresponds to the vierbein field, $D$
is the Lorentz covariant derivative and the small Latin indices take
$a,b,...=0,1,2,3$. Moreover, $l$ is related to the cosmological
constant and defined by $l=\sqrt{3/|\varLambda|}$. Indeed, using
these constraints, the action Eq.(\ref{eq: action sw}) spontaneously
broken down to the Einstein-Cartan action together with a topological
term (more detail see \citep{randono2010}),

\begin{equation}
S_{SW}=-\frac{3}{4\kappa\Lambda}\int\epsilon_{abcd}R^{ab}\wedge R^{cd}-\frac{2}{3}\Lambda\epsilon_{abcd}\left(R^{ab}\wedge e^{c}\wedge e^{d}-\frac{\Lambda}{6}e^{a}\wedge e^{b}\wedge e^{c}\wedge e^{d}\right)\label{eq: action sw 2}
\end{equation}
where the constant is chosen as $\sigma c=-\frac{3}{4\kappa\Lambda}$
and $\kappa=8\pi Gc^{-4}$ is Einstein\textquoteright s gravitational
constant. Here, the first term corresponds to the topological sector
and the remaining terms are the Einstein-Hilbert action with the cosmological
term. Furthermore, we want to note that a similar action to Eq.(\ref{eq: action sw})
can also be found in \citep{Chamseddine1990} where the topological
theories of gravity were investigated in any dimensions. Besides,
to construct an action of this type, it is beneficial to analyze gauged
Wess-Zumino-Witten model and transgression field theory \citep{salgado2014A}. 

Now, we aim to extend $\left(A\right)dS$ gravity by taking into account
the Maxwell symmetry. To do this we start with the generalized $\left(A\right)dS$
algebra,

\begin{equation}
\left[\mathcal{Y}_{AB},\mathcal{Y}_{CD}\right]=i\left(\eta_{AD}\mathcal{Y}_{BC}+\eta_{BC}\mathcal{Y}_{AD}-\eta_{AC}\mathcal{Y}_{BD}-\eta_{BD}\mathcal{Y}_{AC}\right),
\end{equation}
where $\mathcal{Y}_{AB}$ represents the generator of the corresponding
algebra. Moreover, this generator can be decomposed into the following
form, 

\begin{equation}
\mathcal{Y}_{AB}=\frac{1}{2}\left(\mathcal{M}_{AB}+\mathcal{Z}_{AB}\right),
\end{equation}
where $\mathcal{M}_{AB}$ corresponds to a generalized $\left(A\right)dS$
group generator and $\mathcal{Z}_{AB}$ is a new additional antisymmetric
generator responsible for the Maxwell symmetry.. These generators
obey the following Lie algebra,

\begin{eqnarray}
\left[\mathcal{M}_{AB},\mathcal{\mathcal{M}}_{CD}\right] & = & i\left(\eta_{AD}\mathcal{\mathcal{M}}_{BC}+\eta_{BC}\mathcal{\mathcal{M}}_{AD}-\eta_{AC}\mathcal{\mathcal{M}}_{BD}-\eta_{BD}\mathcal{\mathcal{M}}_{AC}\right),\nonumber \\
\left[\mathcal{\mathcal{M}}_{AB},\mathcal{Z}_{CD}\right] & = & i\left(\eta_{AD}\mathcal{Z}_{BC}+\eta_{BC}\mathcal{Z}_{AD}-\eta_{AC}\mathcal{Z}_{BD}-\eta_{BD}\mathcal{Z}_{AC}\right),\nonumber \\
\left[\mathcal{Z}_{AB},\mathcal{Z}_{CD}\right] & = & i\left(\eta_{AD}\mathcal{\mathcal{M}}_{BC}+\eta_{BC}\mathcal{\mathcal{M}}_{AD}-\eta_{AC}\mathcal{\mathcal{M}}_{BD}-\eta_{BD}\mathcal{\mathcal{M}}_{AC}\right).\label{eq: algebra D5 adsMax}
\end{eqnarray}
The resulting algebra is the Maxwell extension of the $\left(A\right)dS$
algebra. To gauge this algebra, We define Lie algebra valued 1-form
$\mathcal{A}\left(x\right)$,

\begin{equation}
\mathcal{A}\left(x\right)=\mathcal{A}^{A}X_{A}=\frac{1}{2}\omega^{AB}\mathcal{M}_{AB}-\frac{1}{2}B^{AB}\mathcal{Z}_{AB},\label{eq: gauge fields-1}
\end{equation}
where $X^{A}=\left(\mathcal{M}_{AB},\mathcal{Z}_{AB}\right)$ correspond
to the generators of the algebra and the associated gauge fields $\mathcal{A}^{A}=\left(\omega^{AB},B^{AB}\right)$
can be described by 1-form fields $\omega^{AB}=\omega_{\mu}^{AB}dx^{\mu}$
and $B^{AB}=B_{\mu}^{AB}dx^{\mu}$, respectively. By the help of corresponding
Lie algebra valued zero-form gauge generator $\zeta(x)$ which is
defined as follows,

\begin{equation}
\zeta\left(x\right)=\zeta^{A}X_{A}=-\frac{1}{2}\tau^{AB}\left(x\right)\mathcal{M}_{AB}-\frac{1}{2}\phi^{AB}\left(x\right)\mathcal{Z}_{AB},\label{eq: gauge generators-1}
\end{equation}
the transformation of connection one-form is given by

\begin{equation}
\delta\mathcal{A}=-d\zeta-i\left[\mathcal{A},\zeta\right],\label{eq: gauge fields var formula}
\end{equation}
and hence we get variations of the gauge fields

\begin{eqnarray}
\delta\omega^{AB} & = & -d\tau^{AB}-\omega_{\,\,\,C}^{[A}\tau^{C|B]}-B_{\,\,\,C}^{[A}\phi^{C|B]},\nonumber \\
\delta B^{AB} & = & -d\phi^{AB}-\omega_{\,\,\,C}^{[A}\phi^{C|B]}+\tau_{\,\,\,C}^{[A}B^{C|B]},\label{eq: var gauge fields}
\end{eqnarray}
where $\tau^{AB}\left(x\right)$ and $\phi^{AB}\left(x\right)$ are
the parameters of the corresponding generators, respectively. The
curvature two-forms are defined to be

\begin{eqnarray}
\mathcal{F}\left(x\right) & =\mathcal{F}^{A}X_{A}= & -\frac{1}{2}\mathcal{R}{}^{AB}\mathcal{M}_{AB}-\frac{1}{2}\mathcal{F}{}^{AB}\mathcal{Z}_{AB},
\end{eqnarray}
where $\mathcal{R}^{AB}$ and $\mathcal{F}^{AB}$ represent the curvatures
corresponding to the respective generators. To find the explicit forms
of these curvatures, we use the following structure equation

\begin{equation}
\mathcal{F}=d\mathcal{A}+\frac{i}{2}\left[\mathcal{A},\mathcal{A}\right],\label{eq: curvature  formula}
\end{equation}
and taking account of the gauge fields in Eq.(\ref{eq: gauge fields-1}),
the group curvatures are found to be,

\begin{eqnarray}
\mathcal{R}^{AB} & = & d\omega^{AB}+\omega_{\,\,C}^{A}\wedge\omega^{CB}+B_{\,\,C}^{A}\wedge B^{CB},\nonumber \\
 & = & R^{AB}+B_{\,\,C}^{A}\wedge B^{CB},\nonumber \\
\mathcal{F}^{AB} & = & dB^{AB}+\omega_{\,\,C}^{[A}\wedge B^{C|B]},
\end{eqnarray}
where $R^{AB}=d\omega^{AB}+\omega_{\,\,C}^{A}\wedge\omega^{CB}$ denotes
the usual $(A)dS$ curvature 2-forms. The transformation properties
of the 2-forms curvature under the infinitesimal gauge transformation
can be derived from the following expression,

\begin{equation}
\delta\mathcal{F}=i\left[\zeta,\mathcal{F}\right],\label{eq: curvature var formula}
\end{equation}
and so one can obtain

\begin{eqnarray}
\delta\mathcal{R}^{AB} & = & \tau{}_{\,\,\,C}^{[A}\mathcal{R}^{C|B]}+\phi_{\,\,\,C}^{[A}\mathcal{F}^{C|B]},\nonumber \\
\delta\mathcal{F}^{AB} & = & \phi_{\,\,\,C}^{[A}\mathcal{R}^{C|B]}+\tau{}_{\,\,\,C}^{[A}\mathcal{F}^{C|B]}.
\end{eqnarray}

Now, we are in a position to construct the gauge-invariant action
under local $\left(A\right)dS$-Maxwell transformations. First, define
the

\begin{eqnarray}
\mathcal{J}^{AB} & = & \mathcal{R}^{AB}+\mathcal{F}^{AB},
\end{eqnarray}
with the following variation 
\begin{equation}
\delta\mathcal{J}^{AB}=\tilde{\tau}{}_{\,\,\,C}^{[A}\mathcal{J}^{C|B]},
\end{equation}
where the shifted zero-form parameter is $\tilde{\tau}^{AB}=\tau^{AB}+\phi^{AB}$.
Moreover, it can be shown that the exterior covariant derivative of
the shifted curvature goes to zero, i.e. $\mathcal{D}\mathcal{J}^{AB}=0$,
where the extended covariant derivative is given by
\begin{equation}
\mathcal{D}\Phi=[d+\tilde{\omega}]\Phi.
\end{equation}
Here, $\tilde{\omega}^{AB}=\omega^{AB}+B^{AB}$ is the shifted connection.
In terms of this connection, the shifted curvature can be written
as
\begin{equation}
\mathcal{J}^{AB}=d\tilde{\omega}^{AB}+\tilde{\omega}_{\,\,C}^{A}\wedge\tilde{\omega}^{CB}.
\end{equation}
Using the shifted curvature 2-form, we can write down the Stelle-West
type action as follows\citep{stelle1979,stelle1980,randono2010}

\begin{equation}
S=-\frac{3}{4\kappa\Lambda c}\int V^{E}\epsilon_{ABCDE}\mathcal{J}^{AB}\wedge\mathcal{J}^{CD}+\alpha\left(c^{2}-V_{A}V^{A}\right).\label{eq: action maxwell sw}
\end{equation}
By varying the action Eq.(\ref{eq: action maxwell sw}) with respect
to $\tilde{\omega}^{AB}$, one obtains

\begin{equation}
\epsilon_{ABCDE}\mathcal{D}V^{E}\wedge\mathcal{J}^{CD}+V^{E}\epsilon_{ABCDE}\mathcal{D}\mathcal{J}^{CD}=0.
\end{equation}
We already know that exterior covariant derivative of the shifted
curvature 2-form is zero. So, the equations of the motion reduce to
\begin{equation}
\epsilon_{ABCDE}\mathcal{D}V^{E}\wedge\mathcal{J}^{CD}=0,\label{eq: eq motion}
\end{equation}
if we impose the constraints similar to Eq.(\ref{eq: constaints}),

\begin{equation}
e^{a}=-l\mathcal{D}V^{a}=-l\tilde{\omega}{}_{\,\,4}^{a},\,\,\,\,\,\,\,\,\mathcal{D}V^{4}=0,\label{eq: constraints maxwell sw}
\end{equation}
where $l=\sqrt{3/|\Lambda|}$, then the action \ref{eq: action maxwell sw}
now takes the following form,

\begin{eqnarray}
S_{SW} & = & -\frac{3}{4\kappa\Lambda}\int\epsilon_{abcd}R^{ab}\wedge R^{cd}-\frac{2}{3}\Lambda\left(\epsilon_{abcd}R^{ab}\wedge e^{c}\wedge e^{d}-\frac{\Lambda}{6}e^{a}\wedge e^{b}\wedge e^{c}\wedge e^{d}\right)\nonumber \\
 &  & +\epsilon_{abcd}\left(2R^{ab}\wedge B_{\,\,e}^{c}\wedge B^{ed}+DB^{ab}\wedge DB^{cd}-\frac{2\varLambda}{3}DB^{ab}\wedge e^{c}\wedge e^{d}\right)\nonumber \\
 &  & +\epsilon_{abcd}\left(2DB^{ab}\wedge B_{\,\,e}^{c}\wedge B^{ed}+B_{\,\,e}^{a}\wedge B^{eb}\wedge B_{\,\,f}^{c}\wedge B^{fd}-\frac{4\varLambda}{3}B_{\,\,e}^{a}\wedge B^{eb}\wedge e^{c}\wedge e^{d}\right).\label{eq: action sw max}
\end{eqnarray}
where $R^{ab}\left(\omega\right)=d\omega^{ab}+\omega_{\,\,c}^{a}\wedge\omega^{cb}$
corresponds the Riemann curvature 2-form. Here, the first term being
Euler topological invariant term does not contribute to the equation
of motion. The second terms correspond to the Einstein-Hilbert action
together with a cosmological term, and the remaining terms contain
the fields $B^{ab}(x)$ coupled to the spin connection and vierbein.
So, we get the Maxwell extension of the Stelle-West action given in
Eq.(\ref{eq: action sw 2}). Moreover, Eq.(\ref{eq: eq motion}) reduces
to the well-known form,
\begin{equation}
\epsilon_{abcd}\mathcal{J}^{ab}\wedge e^{c}=0.
\end{equation}
and this equation leads to an extended version of the Einstein equation
in the coordinate basis as follows,
\begin{equation}
\mathcal{J}_{\mu\nu}-\frac{1}{2}g_{\mu\nu}\mathcal{J}=0.
\end{equation}

\section{Decomposition of $\mathcal{\left(A\right)\text{\ensuremath{d}}S}$-maxwell
algebra}

In the previous section, we briefly reviewed the Stelle-West model
of gravity and gave its extension by using the Maxwell symmetry in
five-dimensions. We also showed that the action Eq.(\ref{eq: action maxwell sw})
reduces to the generalized Einstein-Cartan gravity by choosing special
constraints in Eq.(\ref{eq: constraints maxwell sw}). In this section,
we establish the gauge theory of gravity based on the Maxwell extended
$\left(A\right)dS$ group in 4-dimensional space-time. To do this,
we first decompose the extended $\left(A\right)dS$ algebra in Eq.(\ref{eq: algebra D5 adsMax})
in terms of the following generators,
\begin{equation}
M_{ab}=\mathcal{\mathcal{M}}_{ab},\,\,\,\,\,\,\,\,P_{a}=\sqrt{\frac{\lambda}{2}}\left(\mathcal{\mathcal{M}}_{5a}+\mathcal{Z}_{5a}\right),\,\,\,\,\,\,Z_{ab}=\mathcal{Z}_{ab},\label{eq: decom condition}
\end{equation}
the Lie algebra of the corresponding group is found to be

\begin{eqnarray}
\left[M_{ab},M_{cd}\right] & = & i\left(\eta_{ad}M_{bc}+\eta_{bc}M_{ad}-\eta_{ac}M_{bd}-\eta_{bd}M_{ac}\right),\nonumber \\
\left[M_{ab},Z_{cd}\right] & = & i\left(\eta_{ad}Z_{bc}+\eta_{bc}Z_{ad}-\eta_{ac}Z_{bd}-\eta_{bd}Z_{ac}\right),\nonumber \\
\left[Z_{ab},Z_{cd}\right] & = & i\left(\eta_{ad}M_{bc}+\eta_{bc}M_{ad}-\eta_{ac}M_{bd}-\eta_{bd}M_{ac}\right),\nonumber \\
\left[P_{a},P_{b}\right] & = & i\lambda\left(M_{ab}+Z_{ab}\right),\nonumber \\
\left[M_{ab},P_{d}\right] & = & i\left(\eta_{bd}P_{a}-\eta_{ad}P_{b}\right),\nonumber \\
\left[Z_{ab},P_{d}\right] & = & i\left(\eta_{bd}P_{a}-\eta_{ad}P_{b}\right),\label{eq: ss ads maxwell algebra}
\end{eqnarray}
where the generators $X_{A}=\left\{ P_{a},M_{ab},Z_{ab}\right\} $
correspond to the translations, the Lorentz and the Maxwell transformations.
Here, the constant $\lambda$ has the unit of $L^{-2}$, and it will
be related to the cosmological constant and the metric tensor defined
as $\eta_{ab}=\text{diag}\left(+,-,-,-\right)$. The self-consistency
of this algebra can be checked with the help of the Jacobi identities.
The algebra given by Eq.\ref{eq: ss ads maxwell algebra} is the semisimple
extension of AdS-Maxwell algebra presented in \citep{durka2012}.

We will follow similar methods for establishing a gauge theory of
gravity as in \citep{yang1954,utiyama1956,kibble1961,sciama1964}
by using differential forms. Let us first define the $\left(A\right)dS$-Maxwell
algebra-valued one-form $\mathcal{A}\left(x\right)=\mathcal{A}^{A}X_{A}$
as follows,

\begin{equation}
\mathcal{A}\left(x\right)=e^{a}P_{a}-\frac{1}{2}\omega^{ab}M_{ab}-\frac{1}{2}B^{ab}Z_{ab},\label{eq: gauge fields}
\end{equation}
where $\mathcal{A}^{A}\left(x\right)=\left\{ e^{a},\omega^{ab},B^{ab}\right\} $
are the gauge fields which correspond to the generators of the symmetry
group, respectively. Moreover, the unit dimension of the all gauge
fields have zero other than $[e^{a}]=L$. Here $L$ is considered
as the unit of length. The variation of the gauge field $\mathcal{A}(x)$
under a gauge transformation can be found by using Eq.(\ref{eq: gauge fields var formula})
and the following $\left(A\right)dS$-Maxwell algebra-valued zero-form
gauge generator,

\begin{equation}
\zeta\left(x\right)=y^{a}P_{a}-\frac{1}{2}\tau^{ab}M_{ab}-\frac{1}{2}\phi^{ab}Z_{ab}.\label{eq: gauge generators}
\end{equation}
Thus, one can find the transformation law of gauge fields as follows,
\begin{eqnarray}
\delta e^{a} & = & -dy^{a}-\omega_{\,\,c}^{a}y^{c}-B_{\,\,c}^{a}y^{c}+\tau_{\,\,c}^{a}e^{c}+\phi_{\,\,c}^{a}e^{c},\nonumber \\
\delta\omega^{ab} & = & -d\tau^{ab}-\omega_{\,\,\,c}^{[a}\tau^{c|b]}-B_{\,\,\,c}^{[a}\phi^{c|b]}-\lambda e^{[a}y^{b]},\nonumber \\
\delta B^{ab} & = & -d\phi^{ab}-\omega_{\,\,\,c}^{[a}\phi^{c|b]}+\tau_{\,\,\,c}^{[a}B^{c|b]}-\lambda e^{[a}y^{b]},
\end{eqnarray}
where $y^{a}(x)$, $\tau^{ab}(x)$ and $\phi^{ab}(x)$ are the parameters
of the corresponding generators. The curvature two-forms of the associated
gauge fields $\mathcal{F}\left(x\right)=\mathcal{F}^{A}X_{A}$ are
defined to be

\begin{eqnarray}
\mathcal{F}\left(x\right) & = & \mathcal{F}^{a}P_{a}-\frac{1}{2}\mathcal{R}{}^{ab}M_{ab}-\frac{1}{2}\mathcal{F}{}^{ab}Z_{ab},\label{eq: curvature dec alg}
\end{eqnarray}
where $\mathcal{F}^{a}\left(x\right)$, $\mathcal{R}^{ab}\left(x\right)$
and $\mathcal{F}^{ab}\left(x\right)$ represent the curvatures which
comes from the associated generators. To find the explicit forms of
these curvatures, we have used the structure equation Eq.(\ref{eq: curvature  formula})
together with the gauge fields in Eq.(\ref{eq: gauge fields}). The
group curvature 2-forms are,
\begin{eqnarray}
\mathcal{F}^{a} & = & de^{a}+\omega_{\,\,c}^{a}\wedge e^{c}+B_{\,\,c}^{a}\wedge e^{c},\nonumber \\
\mathcal{R}^{ab} & = & R^{ab}\left(\omega\right)+B_{\,\,c}^{a}\wedge B^{cb}+\lambda e^{a}\wedge e^{b},\nonumber \\
\mathcal{F}^{ab} & = & dB^{ab}+\omega_{\,\,c}^{[a}\wedge B^{c|b]}+\lambda e^{a}\wedge e^{b},
\end{eqnarray}
where $R^{ab}\left(\omega\right)$ denotes the usual Riemann 2-form
tensor. The transformation properties of the curvature 2-forms under
the infinitesimal gauge transformation can be found by using Eq.(\ref{eq: curvature var formula}),
Eq.(\ref{eq: gauge generators}) and Eq.(\ref{eq: curvature dec alg}),
\begin{eqnarray}
\delta\mathcal{F}^{a} & = & -\mathcal{R}_{\,\,c}^{a}y^{c}+\tau_{\,\,c}^{a}\mathcal{F}^{c}-\mathcal{F}_{\,\,c}^{a}y^{c}+\phi_{\,\,c}^{a}\mathcal{F}^{c}\nonumber \\
\delta\mathcal{R}^{ab} & = & \tau{}_{\,\,\,c}^{[a}\mathcal{R}^{c|b]}+\phi_{\,\,\,c}^{[a}\mathcal{F}^{c|b]}+\lambda y^{[a}\mathcal{F}^{b]}\nonumber \\
\delta\mathcal{F}^{ab} & = & \phi_{\,\,\,c}^{[a}\mathcal{R}^{c|b]}+\tau{}_{\,\,\,c}^{[a}\mathcal{F}^{c|b]}+\lambda y^{[a}\mathcal{F}^{b]}
\end{eqnarray}

If we define a shifted curvature as $\mathcal{J}^{ab}=\mathcal{R}^{ab}+\mathcal{F}^{ab}$,
then it transforms as $\delta\mathcal{J}^{ab}=-\mathcal{J}_{\,\,\,c}^{[a}\left(\tau^{c|b]}+\phi^{c|b]}\right)$.
We finally write the MacDowell-Mansouri like \citep{macDowell1977}
action as

\begin{eqnarray}
S & = & \frac{1}{4\kappa\gamma}\int\epsilon_{abcd}\mathcal{J}^{ab}\wedge\mathcal{J}^{cd}\nonumber \\
 & = & \frac{1}{4\kappa\gamma}\int\epsilon_{abcd}R^{ab}\wedge R^{cd}+4\lambda\left(\epsilon_{abcd}R^{ab}\wedge e^{c}\wedge e^{d}+\lambda\epsilon_{abcd}e^{a}\wedge e^{b}\wedge e^{c}\wedge e^{d}\right)\nonumber \\
 &  & +2\epsilon_{abcd}R^{ab}\wedge B_{\,\,e}^{c}\wedge B^{ed}+\epsilon_{abcd}B_{\,\,e}^{a}\wedge B^{eb}\wedge B_{\,\,e}^{c}\wedge B^{ed}\nonumber \\
 &  & +4\lambda\epsilon_{abcd}B_{\,\,e}^{a}\wedge B^{eb}\wedge e^{c}\wedge e^{d}+2\epsilon_{abcd}R^{ab}\wedge DB^{cd}\nonumber \\
 &  & +2\epsilon_{abcd}B_{\,\,e}^{a}\wedge B^{eb}\wedge DB^{cd}+\epsilon_{abcd}DB^{ab}\wedge DB^{cd}+4\lambda\epsilon_{abcd}DB^{ab}\wedge e^{c}\wedge e^{d}\label{eq: action dec msw}
\end{eqnarray}
where $\gamma$ is an arbitrary constant. Furthermore, if we choose
$\lambda=-\frac{\Lambda}{6}$ and $\gamma=-\frac{\Lambda}{3}$ then
the resulting action takes the same form as the action given in Eq.(\ref{eq: action maxwell sw}).

Now, we are in a position to consider the field equations of the theory
and they can be derived from a variational action principle. The equations
of motion can be found by the variation of the action in Eq.(\ref{eq: action dec msw})
with respect to the gauge fields $\omega^{ab}(x)$, $B^{ab}(x)$ and
$e^{a}(x)$, respectively,

\begin{eqnarray}
D\mathcal{J}^{cd}+B_{\,\,\,e}^{[c}\wedge\mathcal{J}^{e|d]} & = & 0,\label{eq: eq of motion 1}\\
\epsilon_{abcd}e^{b}\wedge\mathcal{J}^{cd} & = & 0,\label{eq: eq of motion 2}
\end{eqnarray}
here, we want to note that the variation with respect to $\omega^{ab}(x)$
and $B^{ab}(x)$ lead to the same equation in Eq.(\ref{eq: eq of motion 1}).
Furthermore, one can show that all these equations of motion verify
each other. Making use of he shifted curvature and passing from the
tangent indices to world indices with the help of 

\begin{equation}
e_{a}^{\mu}e_{b}^{\nu}\mathcal{J}^{ab}=\frac{1}{2}\mathcal{J}_{\rho\sigma}^{\mu\nu}dx^{\rho}\wedge dx^{\sigma}
\end{equation}
one writes the field Eq.(\ref{eq: eq of motion 2}) as

\begin{equation}
\mathcal{J}_{\,\,\rho}^{\mu}-\frac{1}{2}\delta_{\,\,\rho}^{\mu}\mathcal{J}=0.
\end{equation}
and expanding the shifted curvature we get,

\begin{eqnarray}
R_{\,\,\rho}^{\mu}\left(\omega\right)-\frac{1}{2}R\left(\omega\right)\delta_{\,\,\rho}^{\mu}+\Lambda\delta_{\,\,\rho}^{\mu} & = & T_{\,\,\rho}^{\mu}\left(B\right)\label{eq: EFE 2}
\end{eqnarray}
where
\begin{equation}
T_{\,\,\rho}^{\mu}\left(B\right)=-\left(e_{a}^{\mu}e_{b}^{\nu}D_{[\rho}B_{\nu]}^{ab}+e_{a}^{\mu}e_{b}^{\nu}B_{[\rho\,\,c}^{a}\wedge B_{\nu]}^{cb}\right)+\frac{1}{2}\delta_{\,\,\rho}^{\mu}\left(e_{a}^{\gamma}e_{b}^{\kappa}D_{[\gamma}B_{\kappa]}^{ab}+e_{a}^{\gamma}e_{b}^{\kappa}B_{[\gamma\,\,c}^{a}\wedge B_{\delta]}^{cb}\right)
\end{equation}
and it represents the tensorial contribution of the Maxwell symmetry.
Therefore, we demonstrated that a new extended framework leads to
the generalized Einstein field equation together with a cosmological
term plus additional energy-momentum tensor as a function of the gauge
field $B^{ab}\left(x\right)$. Moreover, in the limit of $B^{ab}(x)=0$,
Eq.(\ref{eq: EFE 2}) reduces to the well-known Einsteins's gravitational
field equation including the cosmological constant.

\section{Conclusion}

In the present paper, we interested in a gauge theory of gravity based
on the Maxwell extension of $\left(A\right)dS$ algebra. In our extension,
the translation generator satisfies a new commutation relationship
as $\left[P_{a},P_{b}\right]=i\left(M_{ab}+Z_{ab}\right)$. From this
type of extension, we obtained a generalized Lie algebra by unifying
the (anti) de Sitter with the Maxwell algebra presented in Eq.(\ref{eq: algebra D5 adsMax}).
In this generalization, we preserved the semi-simple structure of
de Sitter algebra. We then constructed the gauge theory for the resulting
algebra and establishing a Stelle-West like action and we derived
a generalization of Einstein's field equations. Moreover, we obtained
the semi-simple extension of $\left(A\right)dS$-Maxwell algebra in
Eq.(\ref{eq: ss ads maxwell algebra}) for four-dimensional case by
decomposing the algebra given in Eq.(\ref{eq: algebra D5 adsMax})
under the chosen conditions in Eq.(\ref{eq: decom condition}). After
that, we took this algebra for the construction of gauge theory and
established a MacDowell-Mansouri like action. As a result of these
calculations, we obtained the same field equations as the previous
one in a certain condition. These field equations contain a positive
cosmological constant and additional terms related to the Maxwell
symmetry in addition to Einstein's field equations. So the resulting
gravitational theory can be seen as a generalization of the results
given in \citep{durka2012}. If we take the Maxwell gauge field as
$B^{ab}\left(x\right)=0$, then this gravitational model reduces to
well-known de Sitter gravity. 

We want to remark that it is possible to construct a Yang-Mills like
action \citep{yang1954} based on these extended algebras given in
Eq.(\ref{eq: algebra D5 adsMax}) and Eq.(\ref{eq: ss ads maxwell algebra})
because of their semi-simple characteristics. We know that three of
four fundamental interactions except gravity are Yang-Mills type of
gauge theories. Therefore, the resulting algebras may provide a useful
background to study the unification problem of fundamental interactions
\citep{Blagojevic2002}.

As we mentioned before the Maxwell extended algebras (similar to the
de Sitter algebras) provide a useful background to investigate the
cosmological constant problem. So the unification of de Sitter and
Maxwell algebras may open a new way to analyze the cosmological constant
problem. Furthermore, there may be occurred interesting results if
we construct a cosmological interpretation of the Maxwell extended
theories because, in addition to minimal de Sitter gravity theory,
the Maxwell extended theories contain new energy-momentum tensors
dependent on $B^{ab}\left(x\right)$ fields. Thus, it is known that
additional source term can be related to the dark energy \citep{frieman2008,padmanabhan2009},
the Maxwell fields $B^{ab}\left(x\right)$ may play an important role
to explain the dark energy problem.

\section*{Acknowledgments}

This study is supported by the Scientific and Technological Research
Council of Turkey (TÜB\.{I}TAK) Research project No. 118F364.


\begin{thebibliography}{99}
\bibitem{stelle1979} K. S. Stelle, P. C. West, J. Phys. A: Math.
Gen. \textbf{12}, L205 (1979).

\bibitem{stelle1980} K. S. Stelle, P. C. West, Phys. Rev. D \textbf{21},
1466 (1980).

\bibitem{tseytlin1982} A.A. Tseytlin, Phys. Rev. D \textbf{26}, 3327
(1982).

\bibitem{vervimp1994} T. Verwimp, J. Phys.A: Math. Gen. \textbf{27},
2773 (1994).

\bibitem{townsend1977} P. K. Townsend, Phys. Rev. D \textbf{15},
2795 (1977).

\bibitem{macDowell1977} S. MacDowell, F. Mansouri, Phys. Rev. Lett.
\textbf{38}, 739--42 (1977).

\bibitem{smrz1980} P. K. Smrz, Found. Phys. \textbf{10}, 267 (1980).

\bibitem{Piguet1995} O. Piguet, S.P. Sorella, \textit{Algebraic Renormalization:
Perturbative Renormalization, Symmetries and Anomalies}. Springer
Lect. Notes Phys., vol. M28 (1995).

\bibitem{huang2008} C. G. Huang, H. Q. Zhang, H. Y. Guo, JCAP \textbf{10},
010 (2008).

\bibitem{soroka2005} D. V. Soroka, V. A. Soroka, Phys. Lett. B \textbf{607},
302-305 (2005).

\bibitem{gomis2009} J. Gomis, K. Kamimura, J. Lukierski, JHEP \textbf{08},
39 (2009).

\bibitem{bonanos2009} S. Bonanos, J. Gomis, J. Phys. A: Math. Theor.
\textbf{42}, 145206 (2009).

\bibitem{bonanos2010} S. Bonanos, J. Gomis, K. Kamimura, J. Lukierski,
Phys. Rev. Lett. \textbf{104}, 090401 (2010).

\bibitem{bonanos2010A} S. Bonanos, J. Gomis, K. Kamimura, J. Lukierski,
J. Math. Phys. \textbf{51}, 102301 (2010).

\bibitem{bonanos2010B} S. Bonanos, J. Gomis, J. Phys. A \textbf{43},
015201 (2010).

\bibitem{bacry1970} H. Bacry, P. Combe, J.L. Richard, Nuovo Cimento
\textbf{67}, 267-299 (1970).

\bibitem{schrader1972} R. Schrader, Fortschritte der Physik \textbf{20},
701 (1972).

\bibitem{azcarraga2011} J. A. de Azcárraga, K. Kamimura, J. Lukierski,
Phys. Rev. D \textbf{83}, 124036 (2011).

\bibitem{soroka2012} D. V. Soroka, V. A. Soroka, Phys. Lett. B \textbf{707},
160-162 (2012).

\bibitem{durka2012} R. Durka, Kowalski-Glikman, M. Szczachor, Mod.
Phys. Lett. A \textbf{27}, 1250023 (2012).

\bibitem{Azcarraga2014} J. A. de Azcárraga, J. M. Izquierdo, Nuclear
Phys. B \textbf{885}, 34--45 (2014).

\bibitem{cebecioglu2014} O. Cebecio\u{g}lu, S. Kibaro\u{g}lu, Phys.
Rev. D \textbf{90}, 084053 (2014).

\bibitem{salgado2014} P. Salgado, S. Salgado, Phys. Lett. B \textbf{728},
5-10 (2014).

\bibitem{cebecioglu2015} O. Cebecio\u{g}lu, S. Kibaro\u{g}lu, Phys.
Lett. B \textbf{751}, 131-134 (2015).

\bibitem{concha2015A} P. K. Concha, E. K. Rodriguez, P. Salgado,
JHEP \textbf{08}, 009 (2015). 

\bibitem{penafiel2018} D. M. Peñafiel, L. Ravera, Eur. Phys. J. C
\textbf{78}, 945 (2018).

\bibitem{ravera2018} L. Ravera, Eur. Phys. J. C \textbf{78}, 211
(2018).

\bibitem{kibaroglu2019A} S. Kibaro\u{g}lu, M. Senay, O. Cebecio\u{g}lu,
Mod. Phys. Lett. A \textbf{34}, 1950016 (2019).

\bibitem{kibaroglu2019B} S. Kibaro\u{g}lu, O. Cebecio\u{g}lu, Eur.
Phys. J. C. \textbf{79}, 898 (2019).

\bibitem{kibaroglu2020A} S. Kibaro\u{g}lu, O. Cebecio\u{g}lu, Phys.
Lett. B \textbf{803}, 135295 (2020).

\bibitem{concha2019A} P. Concha, L. Ravera, E. Rodriguez, JHEP \textbf{01},
192 (2019).

\bibitem{concha2019B} P. Concha, Phys. Lett. B \textbf{792}, 290-297
(2019).

\bibitem{freund1974} P. G. O. Freund, Ann. Phys. \textbf{84}, 440
(1974). 

\bibitem{bramson1974} B. D. Bramson, Phys. Lett. A \textbf{47}, 431
(1974).

\bibitem{fedoruk2012} S. Fedoruk, J. Lukierski, Phys. Lett. B \textbf{718},
646 (2012).

\bibitem{fedoruk2013} S. Fedoruk, J. Lukierski, JHEP \textbf{02},
128 (2013).

\bibitem{fedoruk2013A} S. Fedoruk, J. Lukierski, J. Phys. Conf. Ser.
\textbf{474}, 012016 (2013).

\bibitem{hoseinzadeh2015} S. Hoseinzadeh, A. Rezaei-Aghdam, Eur.
Phys. J. C \textbf{75}, 227 (2015).

\bibitem{leclerc2006} M. Leclerc, Annals of Physics \textbf{321},
708--743 (2006).

\bibitem{randono2010} A. Randono, Class. Quantum Grav. \textbf{27},
215019 (2010). 

\bibitem{Chamseddine1990} A. H. Chamseddine, Nucl. Phys. B \textbf{346},
213 (1990).

\bibitem{salgado2014A} P. Salgado, R. J. Szabo, O. Valdivia, Phys.
Rev. D \textbf{89}, 084077 (2014).

\bibitem{yang1954} C. N. Yang, R. L. Mills, Phys. Rev. \textbf{96},
191-195 (1954).

\bibitem{utiyama1956} R. Utiyama, Phys. Rev. \textbf{101}, 1597 (1956). 

\bibitem{kibble1961} T. W. B. Kibble, J. Math. Phys. \textbf{2},
212 (1961). 

\bibitem{sciama1964} D. W. Sciama, Rev. Mod. Phys. \textbf{36}, 463
(1964).

\bibitem{sobreiro2012} R. F. Sobreiro, A. A. Tomaz, V. J. Vasquez
Otoya, Eur. Phys. J. C \textbf{72}, 1991 (2012).

\bibitem{Blagojevic2002} M. Blagojevi\'{c}, Gravitation and Gauge
Symmetries (IOP, Bristol, 2002).

\bibitem{frieman2008} J. Frieman, M. Turner and D. Huterer, Annu.
Rev. Astron. Astrophys. \textbf{46}, 385 (2008).

\bibitem{padmanabhan2009} T. Padmanabhan, Adv. Sci. Lett. \textbf{2},
174 (2009).
\end{thebibliography}
\end{document}